# Composite vortex-ring solitons in Bessel photonic lattices

Yaroslav V. Kartashov* and Lluis Torner

*ICFO-Institut de Ciencies Fotoniques, and Department of Signal Theory and Communications, Universitat Politecnica de Catalunya, 08034 Barcelona, Spain*

Victor A. Vysloukh

*Departamento de Fisica y Matematicas, Universidad de las Americas – Puebla, Santa Catarina Martir, 72820 Puebla, Mexico*

We reveal the existence of dynamically stable composite topological solitons in Bessel photonic lattices imprinted in focusing Kerr-type nonlinear media. The new stable composite solitons are made of a vortex-ring with unit topological charge incoherently coupled to a fundamental soliton. The stabilization of the otherwise highly azimuthally unstable vortex-rings is provided under suitable conditions, which we study in detail, by cross-phase-modulation coupling with the fundamental soliton companion.

*OCIS codes: 190.5530, 190.4360, 060.1810*

Bright vortex solitons or vortex-rings, which contain a phase singularity nested in a ring shaped beam and that carry orbital angular momentum, appear in many systems modeled by focusing nonlinearities. However in homogeneous self-focusing media they typically undergo symmetry-breaking instabilities and self-fragment themselves into sets of fundamental solitons which fly off the initial ring profile.[1-11] Stabilization of such beams has been predicted theoretically in the models with competing cubic-quintic or quadratic-cubic nonlinearities (see Refs. [12-15] and references therein). Stabilization via nonlinearity management was proposed recently in Bose-Einstein condensates.[16] Ref. [17] suggests a novel scheme for stabilization of vortex solitons by placing them on a defect in the photonic crystal fiber. Two additional different approaches that have been put forward in the search of stabilization are the incoherent coupling of a vortex-ring with a bright soliton via a cross-phase-modulation thus forming composite solitons in saturable focusing media,[18-21] and the utilization of photonic lattices.[22-25] In this paper we discover a new possibility that is possible because of the ingredients that come from both these approaches.

Spatial modulation of the linear refractive index profoundly affects the diffraction properties of the beams and thus the properties of solitons, including the vortex ones. Lattice solitons were proposed theoretically and experimentally demonstrated in photorefractive crystals where modulation of refractive index can be created optically.[26-33] If refractive index modulation is deep enough, as a consequence creating an array of evanescently coupled waveguides, the formation of discrete solitons strongly localized in neighboring array sites is possible.[34,35] Recently experimental observation of discrete vortex solitons in optically induced photonic lattices was reported[22,23] and the strong stabilizing action of the lattice was confirmed.[24] To date the main stream of activity was related to periodical optical lattices with a rectangular, or honeycomb symmetry, that are induced by several interfering plane waves. However, Bessel photonic lattices with the cylindrical symmetry are very attractive because they introduce a new cylindrical symmetry, and can be induced by corresponding non-diffracting beams.[25] Recently we reported salient properties of single-spot and dipole-mode solitons in such lattices imprinted in focusing media and presented reach variety of rotating soliton interactions.[36,37]

In this paper we introduce the new family of bright composite vortex solitons in Bessel photonic lattices imprinted in focusing Kerr-type medium. We found that even in the absence of nonlinearity saturation, the cross-phase-modulation coupling with stable vorticityless component stabilizes vortex soliton component with unit topological charge that is azimuthally unstable when propagating alone in Bessel lattice with focusing nonlinearity. Such suppression of the azimuthal modulational instability of composite vortex solitons serves as an illuminating example of Bessel photonic lattices applications. Notice, that in contrast to vortex solitons supported by periodic lattices whose stabilization is achieved because of the strong azimuthal modulation of vortex intensity profile,[22-24] the stable composite vortex solitons in Bessel lattices preserve ideal ring-like shape.

We address the propagation of two incoherently coupled laser beams along the $z$ axis in a bulk medium with the focusing cubic nonlinearity and transverse modulation of refractive index described by the system of coupled nonlinear Schrödinger equations for dimensionless light field amplitudes $q_{1,2}$:

$$i\frac{\partial q_1}{\partial \xi} = -\frac{1}{2}\left(\frac{\partial^2 q_1}{\partial \eta^2} + \frac{\partial^2 q_1}{\partial \zeta^2}\right) - q_1(|q_1|^2 + |q_2|^2) - pR(\eta,\zeta)q_1,$$
$$i\frac{\partial q_2}{\partial \xi} = -\frac{1}{2}\left(\frac{\partial^2 q_2}{\partial \eta^2} + \frac{\partial^2 q_2}{\partial \zeta^2}\right) - q_2(|q_1|^2 + |q_2|^2) - pR(\eta,\zeta)q_2.$$
(1)

Here longitudinal $\xi$ and transverse $\eta,\zeta$ coordinates are scaled to the diffraction length $L_{\mathrm{d}} = kr_0^2$ and input beam width $r_0$, respectively, where $k = n\omega/c$ is the wave number, $n$ is the refractive index. The parameter $p = \delta n \omega L_{\mathrm{d}}/c$ is proportional to the depth $\delta n$ of the refractive index modulation, and the function $R(\eta,\zeta) = J_0^2[(2b_{\mathrm{lin}})^{1/2} r]$ with $r^2 = \eta^2 + \zeta^2$ stands for the transverse profile of refractive index; the parameter $b_{\mathrm{lin}}$ is related to the radiuses of the rings in the zero-order Bessel lattice.

It should be pointed out that the function $q(\eta,\zeta,\xi) = J_0[(2b_{\mathrm{lin}})^{1/2} r]\exp(-ib_{\mathrm{lin}}\xi)$ is an exact solution of linear homogeneous Eq. (1) (i.e., with $p = 0$) and thus describes a nondiffracting zero-order Bessel beam. Such kind of beams can be formed experimentally by means of axicones or computer generated holograms. We suppose that the lattice refractive index profile depends on the intensity of the zero-order Bessel beam as it occurs in photorefractive crystals and Bose-Einstein condensates (Fig. 1(a)). The depth of refractive index modulation in the lattice is assumed to be small compared with the unperturbed index and is of the order of nonlinear contribution to the refractive index. If vector soliton components launched into a slow Kerr-type medium share the same polarization, which is orthogonal to polarization of lattice-creating beam, they will fill the same lattice formed by the nondiffracting Bessel beam. Notice that lattice-creating beam should experience almost no nonlinearity-induced phase shift, while soliton beams should propagate in essentially nonlinear regime that requires high anisotropy of nonlinearity (similarly to photorefractive crystals). Among the conserved quantities of Eq. (1) are the total and partial energy flows:

$$U = U_1 + U_2 = \int_{-\infty}^{\infty}\int_{-\infty}^{\infty}(|q_1|^2 + |q_2^2|)d\eta d\zeta.$$
(2)

We search for solutions of Eq. (1) with invariable radial profile and helical phase distribution in the form $q_1(\eta,\zeta,\xi) = w_1(r)\exp(im_1\phi)\exp(ib_1\xi)$ and $q_2(\eta,\zeta,\xi) = w_2(r)\exp(im_2\phi)\exp(ib_2\xi)$, where $\phi$ is the azimuth angle, $w_{1,2}(r)$ are real functions, $m_{1,2}$ are topological charges, and $b_{1,2}$ are propagation constants. Substitution of this light field into Eq. (1) yields the system of equations

$$\frac{d^2 w_{1,2}}{dr^2} + \frac{1}{r}\frac{dw_{1,2}}{dr} - \frac{m_{1,2}^2 w_{1,2}}{r^2} - 2b_{1,2}w_{1,2} + 2w_{1,2}(w_1^2 + w_2^2) + 2pRw_{1,2} = 0, \qquad (3)$$

that was solved numerically with a relaxation method to find the soliton profiles. Mathematically, lattice soliton families are defined by the propagation constants $b_{1,2}$, the topological charges $m_{1,2}$, the lattice depth $p$, and the $b_{\text{lin}}$ parameter. Since one can use scaling transformations $q_{1,2}(\eta,\zeta,\xi,p) \to \chi q_{1,2}(\chi\eta,\chi\zeta,\chi^2\xi,\chi^2 p)$ to obtain various families of solitons from a given one, we set the transverse scale in such way that $b_{\text{lin}} = 2$ and varied $b_{1,2}$ and $p$. To analyze soliton stability we searched for perturbed solutions $q_{1,2} = [w_{1,2}(r) + u_{1,2}(r,\xi)\exp(in\phi) + v_{1,2}^*(r,\xi)\exp(-in\phi)]\exp(ib_{1,2}\xi + im_{1,2}\phi)$ of Eq. (1), where the perturbation components $u_{1,2}, v_{1,2}$ could grow with the complex rate $\delta$ upon propagation, and $n$ is the azimuthal index of the perturbation. Linearization of Eq. (1) in the vicinity of a stationary solution yields an eigenvalue problem that was solved numerically.

In the scalar case ($q_2 \equiv 0$) Eq. (1) admits fundamental ($m_1 = 0$) soliton solutions that are stable in the entire domain of their existence for deep enough lattices, and vortex ($m_1 > 0$) soliton solutions that are oscillatory unstable. Both fundamental and vortex solitons become spatially extended and cover many lattice rings as the propagation constant approaches the cut-off value for soliton existence. Typically the cut-off for vortex solitons is lower than that for fundamental soliton. At high energy flows solitons of both types are localized primarily in the central guiding core of Bessel lattice.

We have found a rich variety of composite lattice solitons with different combinations of topological charges $m_{1,2}$ but further we concentrate on the only stable combination corresponding to $m_1 = 1$ and $m_2 = 0$. Typical radial profile of such composite soliton is presented in Fig. 1(b). The existence domain of the composite soliton state in $b_1$ broadens with growth of $b_2$. For a fixed $b_2$ vortex component disappears at $b_1 = b_{\text{low}}$, while at $b_1 = b_{\text{up}} > b_{\text{low}}$ fundamental component vanishes and one gets scalar vortex soliton. This feature is illustrated in Fig. 1(d) that shows energy sharing between soliton constituents $S_{1,2} = U_{1,2}/U$ versus $b_1$. Total energy flow is a monotonically growing function of $b_1$ at fixed $b_2$. It should be mentioned that lower boundary for existence of composite solitons on $b_2$ is given roughly by the cut-off for propagation constant of fundamental solitons supported by zero-order Bessel lattice. This value monotonically increases with growth of the lattice depth $p$. Since lower amplitudes are necessary to support soliton propagation in the presence of lattice, amplitudes $w_{1,2}$ decrease with growth of $p$ at fixed total energy flow, while radial oscillations on soliton profile due to the refractive index modulation become more pronounced.

Since the second vorticityless component would be linearly stable in Bessel lattice in the absence of the first vortex component, it is expectable that second component might have strong stabilizing action on the first component via the cross-phase modulation, at least near the lower cut-off $b_{\text{low}}$ for existence of composite soliton states, where $w_1$ is small. We have found that this is the case by solving the linear eigenvalue problem for perturbations $u_{1,2}, v_{1,2}$. The real part of growth rate can be nonzero only for perturbations with azimuthal indexes $n = 0,1,2$. Typically major part of the instability domain for composite solitons is associated with perturbations carrying topological charge $n = 1$ (see Fig. 2(a) and 2(b)), while $\text{Re}(\delta)$ for perturbation with $n = 2$ becomes nonzero closer to the upper cut-off $b_{\text{up}}$. Growth rate is complex for vortex type perturbations ($n = 1,2$) and real for vorticityless ones ($n = 0$). Perturbation component $u_1$ in the presence of lattice is weakly localized and exhibit slow decaying oscillations as $r \to \infty$ (Fig. 2(b)).

The central result of this paper is that composite vortex solitons are free from instabilities in a considerable part of their existence domain close to the lower cut-off on $b_1$ (for $b_{\text{low}} \leq b_1 < b_{\text{cr}}$). Bessel photonic lattice imprinted in a bulk Kerr-type medium stabilizes vorticityless component that, in turn, via cross-modulation resists to the development of the azimuthal instability of the vortex component. In other words, the presence of the robust and self-restoring vorticityless component is crucial to provide a dumping feedback for azimuthal perturbations. For a fixed lattice depth the stability domain broadens with growth of the amplitude of vorticityless component (or, equivalently, with growth of $b_2$, Fig. 2(c)). It is important to note that close to the critical for stabilization value of the propagation constant $b_{\mathrm{cr}}$ the energy flow carried by the vortex component of stable composite soliton can exceed that carried by vorticityless component. In this case vortex component cannot be considered just as linear mode of the waveguide created by the lattice and the relatively intensive vorticityless component. The influence of the lattice depth on stability of composite solitons at fixed $b_2$ is illustrated in Fig. 2(d). In homogeneous Kerr-type medium ($p \to 0$) vector solitons are unstable in the entire domain of their existence, but with growth of the lattice depth the stability domain broadens, and then shrinks again as $p$ approaches a threshold value for existence of vector solitons at given $b_2$.

To confirm results of linear stability analysis we integrated numerically Eq. (1) with the boundary conditions $q_{1,2}(r,\phi,\xi=0) = w_{1,2}(r)\exp(im_{1,2}\phi)[1+\rho_{1,2}(r,\phi)]$, where $\rho_{1,2}(r,\phi)$ describe broadband noise with the variance $\sigma_{\mathrm{noise}}^2$. We have found that break up of unstable composite solitons in the lattice is accompanied by intense radiation and growing oscillations of the amplitude of vortex component (Fig. 3(a)). Asymptotically it results in formation of simplest vector soliton, whose components have no vorticity and feature the same functional profiles with maximums located at $r=0$. For stable vector solitons direct simulations of Eq. (1) entirely confirm results of linear stability analysis. In the presence of noise stable solitons conserve their input structure for hundreds of the diffraction lengths as shown in Fig. 3(b).

We thus conclude that the analysis of the properties of composite vortex solitons in Bessel photonic lattices reveals that the cross-phase-modulation incoherent coupling with stable vorticityless component leads to stabilization of the otherwise unstable vortex soliton with unit topological charge. Results reported here are expected to hold for more general settings, e.g. more complicated lattices with radial symmetry, and higher values of cross-phase-modulation strength. We stress that the predicted existence of stable composite vortex solitons, which do not self-destroy by azimuthal modulational instabilities, is another important example of the unique phenomena afforded by Bessel photonic lattices.

*On leave from Physics Department, M. V. Lomonosov Moscow State University, Russia. This work has been partially supported by the Generalitat de Catalunya, by the Ramon y Cajal program and by the Spanish Government through grant BFM2002-2861.

# Figure captions

Figure 1. (a) Zero-order Bessel photonic lattice. (b) Soliton profile corresponding to point marked by circle in dispersion diagram (c). (d) Energy sharing versus propagation constant $b_1$ at $b_2 = 7$. Lattice depth $p = 8$.

Figure 2. (a) Real part of the growth rate of perturbation with azimuthal index $n = 1$ at $b_2 = 7$ and $p = 8$. (b) Real parts of perturbation components corresponding to $n = 1$, $b_1 = 2$, $b_2 = 7$, $p = 8$. (c) Areas of stability and instability (shaded) on $(b_1, b_2)$ plane at $p = 8$. (d) Areas of stability and instability (shaded) on $(b_1, p)$ plane at $b_2 = 7$.

Figure 3. (a) Decay of the unstable vector soliton at $b_1 = 2$, and (b) its stable propagation at $b_1 = 1.2$ in the presence of white input noise with the variance $\sigma_{\text{noise}}^2 = 0.01$. Intensity distribution of first component is shown at different propagation distances. Lattice depth $p = 8$ and propagation constant $b_2 = 7$.

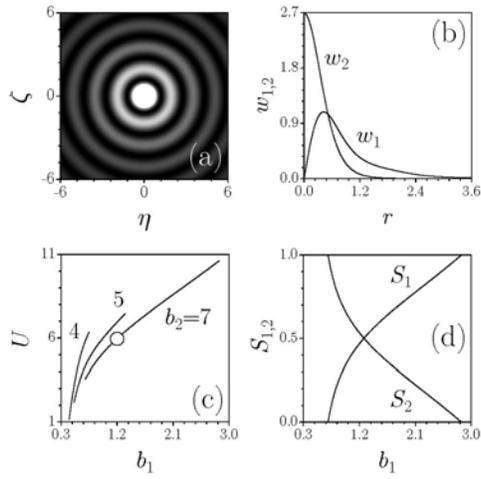

Figure 1.  (a) Zero-order Bessel photonic lattice. (b) Soliton profile corresponding to point marked by circle in dispersion diagram (c). (d) Energy sharing versus propagation constant $b_1$ at $b_2 = 7$. Lattice depth $p = 8$.

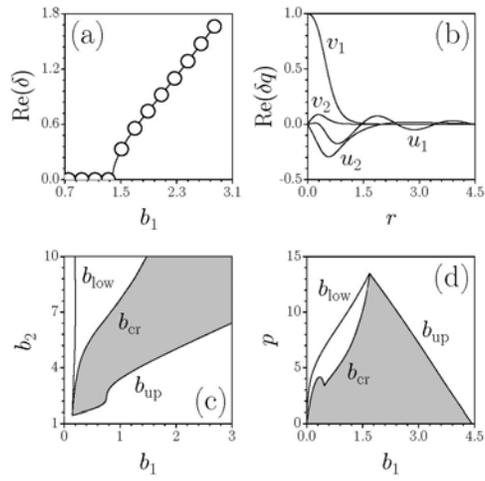

Figure 2. (a) Real part of the growth rate of perturbation with azimuthal index $n=1$ at $b_2=7$ and $p=8$. (b) Real parts of perturbation components corresponding to $n=1$, $b_1=2$, $b_2=7$, $p=8$. (c) Areas of stability and instability (shaded) on $(b_1,b_2)$ plane at $p=8$. (d) Areas of stability and instability (shaded) on $(b_1,p)$ plane at $b_2=7$.

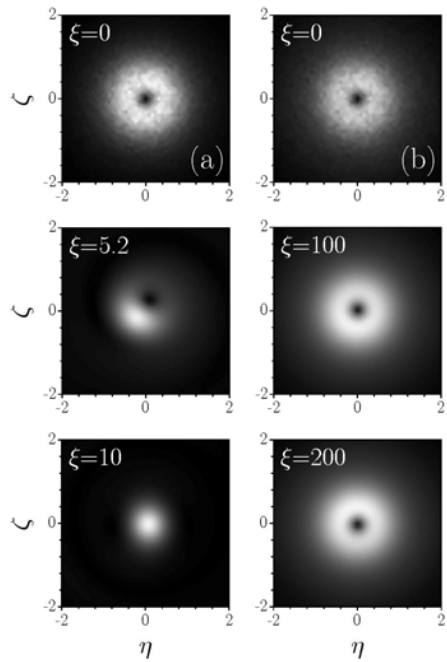

Figure 3.  (a) Decay of the unstable vector soliton at $b_1 = 2$, and (b) its stable propagation at $b_1 = 1.2$ in the presence of white input noise with the variance $\sigma^2_{\text{noise}} = 0.01$. Intensity distribution of first component is shown at different propagation distances. Lattice depth $p = 8$ and propagation constant $b_2 = 7$.